\documentclass[aps,prd,eqsecnum,nofootinbib,twocolumn,floatfix]{revtex4}
\usepackage{graphicx}
\usepackage{times,mathptm}
\newcommand{\beq}{\begin{equation}}
\newcommand{\eeq}{\end{equation}}
\newcommand{\bea}{\begin{eqnarray}}
\newcommand{\eea}{\end{eqnarray}}

\newcommand{\D}{\Delta}

\renewcommand{\l}{\lambda}

\renewcommand{\b}{\beta}

\newcommand{\p}{\partial}

\newcommand{\g}{\gamma}
\newcommand{\n}{\nu}
\newcommand{\m}{\mu}

\newcommand{\s}{\sigma}

\renewcommand{\th}{\theta}
\renewcommand{\to}{\rightarrow}

\newcommand{\e}{\epsilon}

\renewcommand{\t}{\tau}
\newcommand{\rf}[1]{(\ref{#1})}

\begin{document}
%
%
\title{Equation of State of Gluon Plasma from Fundamental Modular Region}
\author{Daniel Zwanziger}
\affiliation{Physics Department, New York University, New York, NY~10003, USA}
\email{daniel.zwanziger@nyu.edu}
\date{\today}
\begin{abstract}

Despite considerable practical success in dealing with the gluon plasma, finite-temperature perturbation theory suffers at the fundamental level  from infrared divergences discovered by Linde.  However if gauge or Gribov copies are properly eliminated from the physical state space, infrared modes are strongly suppressed.  We describe the gluon plasma in zeroth order as a gas of free quasi-particles with a temperature-independent dispersion relation of Gribov type, 
$E(k) = \sqrt{k^2 + {M^4 \over k^2}}$, that results from the reduction of the physical state space.  
The effective mass ${M^2 \over k}$ controls infrared divergences and allows finite calculable corrections.  The equation of state of this gas is calculated and compared with numerical lattice data.

\end{abstract}
\maketitle
%
%

\section{Introduction}

The equation of state (EOS) of the quark-gluon plasma is of vital interest at present because of the exciting RHIC experiments at Brookhaven National Laboratory~\cite{star}.  Here we shall be concerned with the pure gluon plasma, leaving quarks for another occasion.  The EOS of the gluon plasma is rather precisely known from numerical studies in lattice gauge theory~\cite{Karsch96}, and there exist excellent phenomenological fits to the EOS of the gluon plasma in the deconfined phase~\cite{Engels89}.   In contrast with this practical success, it was discovered by Linde~\cite{Linde} as far back as 1980 that finite-temperature perturbative QCD suffers at a fundamental level from infrared divergences, which suggests that finite-temperature perturbation theory neglects an essential feature of QCD.  It has been proposed~\cite{Pisarski81} to control these divergences by introducing a magnetic mass $m \sim g^2T$.  It has also been proposed~\cite{ZZ} that the divergences stem from an inadequate application of the principle of gauge equivalence.    Indeed in 1978, two years before Linde's discovery of the infrared divergence~\cite{Linde}, Gribov showed that infrared modes are strongly suppressed when gauge equivalence is imposed at the non-perturbative level~\cite{Gribov}.  

According to the principle of gauge equivalence, two different configurations that are related by a gauge transformation,
$A_2 = {^g}A_1$, represent the same physical state.  The gauge transformation,
${^g}A_i = g^{-1} A_i g + g^{-1}\partial_i g$,
is highly non-linear,  and the physical state space (which is the quotient of the space of connections modulo gauge transformations) is not a linear space.  This geometric property of gauge theory shows up in practice when one fixes a gauge, which is simply a coordinatization of the physical state space.  Two or more different configurations may be gauge equivalent even though both satisfy the linear Coulomb gauge condition, 
$\sum_{i = 1}^3\partial_i A_i = 0$. 
When enumerating physical states, only one of these ``Gribov" or gauge copies should be counted, so the space of physical states is {\it reduced} to the fundamental modular region (FMR), a region that is free of Gribov copies.  Gribov~\cite{Gribov} found that the dispersion relation $E(k) = k$ gets modified because of the reduction of the physical state space, and he obtained instead
\beq
\label{dispersion}
E(k) = \sqrt{ k^2 + { M^4 \over k^2 } },
\eeq 
where $k = |{\bf k}|$, and $M$ is a QCD mass scale.  The reduction of the physical state space was originally proposed as an essential feature of the confinement mechanism~\cite{Gribov, Feynman81}.  However statistical mechanics is primarily a matter of counting states, and the reduction of the physical state space required by the gauge principle influences the EOS at all temperatures.  Here we shall be concerned with its effect in the deconfined phase.   

	 It  is known from numerical studies~\cite{Karsch96} that at high temperature the EOS of the gluon plasma approaches the Stefan-Boltzmann law, $\epsilon = 3p = 3 c_{SB} T^4$, where $\epsilon$ is the energy per unit volume, $p$ is the pressure, $T$ is the temperature, and 
$c_{\rm SB} = { \pi^2 \over 45} (N^2 - 1)$ in SU(N) gauge theory.  Thus it seems reasonable to describe the gluon plasma at high temperature in first approximation as a gas of non-interacting quasi-particles.  We shall describe the quasi-particles by the Gribov dispersion relation \rf{dispersion} or a similar one, for $E(k)$ is only approximately known.  Fortunately the results obtained hold under rather general conditions on $E(k)$.  We call the gas of non-interacting quasi-particles, with modified dispersion relation, the FMR gas.

	The FMR gas closely resembles early phenomenological models of the gluon plasma~\cite{DeGrand87}, characterized by a temperature-independent infrared cut-off $K$, because the effective mass ${M^2 \over k}$, which is large at low $k$,
provides an infrared cut-off at $k \sim M$.  Those models, which were developed approximately a decade after Gribov's paper, were inspired by the idea that low-momentum gluons are confined but high-momentum gluons are effectively free, and they successfully capture the gross features of the EOS of the gluon plasma.  Phenomenological models were
subsequently improved ~\cite{Engels89}, by making the cut-off $K(T)$ temperature dependent and in other ways, to give a good fit in the transition region, at the cost of additional parameters.  The FMR gas is not intended to compete in precision with these improved  phenomenological models.  Rather it is conceived as the zeroth order approximation that allows {\it calculable} corrections by an iterative procedure, for example of Dyson-Schwinger type (presently being developed).  
Successive iterations yield perturbative-type corrections to the FMR gas that are expected to give good results at high~$T$.
They are similar to thermal perturbation theory
~\cite{Kapusta, Andersen}, but the infrared divergences of thermal perturbation theory~\cite{Linde}, that may be controlled by introducing a magnetic mass $m \sim g^2 T$~\cite{Pisarski81}, are here controlled by the effective mass~${M^2 \over k}$. 
It has been noted by Rischke et al~\cite{DeGrand87} that
infrared divergences are also controlled when the cut-off model is iterated, and that non-analytic corrections of order $g^3$ are absent.

	Many aspects of the Gribov scenario have been verified in recent investigations.  Infrared suppression of the gluon propagator in Coulomb gauge has been observed in numerical simulation  \cite{CZ}, but less strongly in \cite{Langfeld}, and in Landau gauge in 3-dimensions \cite{Cucchieri03}.  It has also been found in variational calculations in Coulomb gauge  \cite{Szczepaniak}, and in  Schwinger-Dyson calculations in Coulomb  \cite{Zwanziger03}  and in Landau \cite{Smekal} gauge.  A long-range color-Coulomb potential was found in numerical simulations in Coulomb gauge in the deconfined phase \cite{Us}. 
 
	The questions we wish to address here are:  (i) What is the equation of state of the FMR gas?  (ii)  How does it compare with the EOS that is known from numerical studies?

\section{EOS of the FMR gas}

	In the Stefan-Boltzmann limit, the degrees of freedom for each gluon momentum ${\bf k}$ are the two states of polarization, each with color multiplicity $(N^2-1)$.  These are precisely the degrees of freedom in Coulomb gauge, and we shall use this gauge for our calculation.  It is a ``physical" gauge without negative metric states, and all constraints are satisfied identically.  Although the Coulomb gauge is not manifestly Lorentz covariant, this is not necessarily a disadvantage in the deconfined phase because at finite $T$
the heat bath provides a preferred Lorentz frame, and the manifest symmetries of the Coulomb gauge are the symmetries of the physical problem at hand. 

The partition function of the FMR gas is given by the Planck distribution,
\beq
\label{partition}
Z = \prod_n \ [ \ 1 - \exp(- \beta E_n) \ ]^{-1},
\eeq
where $\beta = 1/T$ is the inverse temperature, and
$n = ({\bf k},\l,a)$, where ${\bf k}$ is 3-momentum, $\l = 1, 2$ 
is polarization, and $a = 1, ... (N^2 - 1)$ is
color.   With 
$\sum_{{\bf k},\l,a} \rightarrow  
(N^2 -1) V  \pi^{-2} \int dk k^2$,
where $V$ is the 3-volume, this gives
\beq
\label{logz}
\ln Z =  -  (N^2 -1) V  \pi^{-2} \int_0^\infty dk k^2 
\ \ln \{  \ 1 - \exp[-\beta E(k) ] \ \}  ,
\eeq
and, with energy density
$\e = - { 1 \over V} { { \p \ln Z} \over {\p \beta} }$, we have 
\beq
\label{energyx}
\e =  { { (N^2-1) } \over \pi^2 } 
\int_0^\infty dk \ { { k^2 } \over {\exp[\beta E(k)] - 1 } } \ E(k) .
\eeq
For a homogeneous system, the pressure is given by
$p = T{\p \ln Z \over \p V}  =  {T \over V} \ln Z = -f$, 
and the entropy density by $Ts = \e + p$,
where $f = F/V$ is the free energy per unit volume.
After an integration by parts, one has
\beq
\label{pressurex}
p =   { { (N^2-1) } \over {3\pi^2} }
\int_0^\infty dk \ { k^3 \over {\exp[\beta E(k)] - 1 } } \ 
{ { \p E(k) } \over {\p k} }, 
\eeq
From the last two equations, we obtain for the trace anomaly $\th \equiv \e - 3 p = \langle G^2\rangle_0 - \langle G^2\rangle_T$, 
\beq
\label{anomalyx}
\th =   { { (N^2-1) } \over {\pi^2} }
\int_0^\infty dk \ { k^4 \over {\exp[\b E(k)] - 1 } } \ 
{  \p   \over \p k } \Big({  - E(k)  \over k }\Big), 
\eeq
where $G^2 \equiv (2g^3)^{-1}(-\b)G_{\m\n}^a G^{\m\n a}$ is the gluon condensate~\cite{Leutwyler}.  The integrand is positive when
${  \p   \over \p k } \Big({  E(k)  \over k }\Big) < 0$, which is
a sufficient condition for $\e(T) > 3p(T)$. 
We also have $\e = T^2 { \p \over \p T} ( { p \over  T } )$,  from which it follows that
the trace anomaly may be written
$\th = \e - 3p = T^5 { \p \over \p T} ( { p \over  T^4 } ).$
Upon integration this yields
\beq
\label{peint}
p = c_{\rm SB}T^4  - T^4 \int _T^\infty dT' \ T'^{-5} \ \th(T').
\eeq


\section{FMR gas at high temperature}

Suppose that the leading deviation of $E(k)$ from $k$ at high $k$ is expressed by a power law, 
\begin{equation}
E(k) /k = 1 + c / k^\g + ... \ \ \ .
\end{equation}
For the Gribov dispersion relation $E(k) = \sqrt{k^2 + M^4/k^2}$ one has $\g = 4$, whereas for a gluon mass, 
$E(k) = \sqrt{k^2 + m^2}$, one has $\g = 2$.  The gluon condensate has dimension 4, which leads one to expect $\g = 4$, whereas if there were a condensate of dimension 2, one would expect $\g = 2$.  The asymptotic behavior of the EOS is qualitatively different for $\g$ greater or less than~3.  We suppose 
$\g > 3$, and consequently the deviation from the ultraviolet behavior $E(k) = k$ is soft.
 
	For $\g > 3$, the asymptotic high-$T$ limit of the trace anomaly is obtained from the substitution
$ \exp[{ E(k) \over T }] - 1 \to { E(k) \over T }$ in \rf{anomalyx}.  This substitution cannot be made in the integrals for $p$ and $\e$ because they would diverge.  It gives a linear asymptotic trace anomaly,
\begin{equation}
\label{lata}
\th  =  L \ T  + O(1),
\end{equation}
where
\bea
\label{sumrule}
L  & = &   { { (N^2-1) } \over {\pi^2} }
\int_0^\infty dk \ { k^4 \over E(k) } \ 
{  \p   \over \p k } \Big({  - E(k)  \over k }\Big)       \cr
  & = & 3 (N^2-1) \pi^{-2}
\int_0^\infty dk \ k^2  \  \ln [  E(k) / k ], 
 \eea 
is an integral that converges for $\g > 3$ (as we have supposed), and is positive for $E(k) > k$.  These are sufficient conditions for the linear asymptotic form \rf{lata} with $L > 0$.   
Although $L$ describes the high-$T$ limit of the FMR gas, the last integral involves all momenta~$k$.  
For the special case of the Gribov dispersion relation \rf{dispersion}, one obtains  
\begin{equation}
L = (N^2 -1) \ (\pi \sqrt{2})^{-1}  \ M^3,
\end{equation}
which is proportional to $M^3$ although only $M^4$ appears in the dispersion relation. 

The pressure at high $T$ is obtained from \rf{peint} which yields
\beq
\label{asymptp}
p = c_{SB} \ T^4 - (1/3) L \ T + O(1),
\eeq
Thus for the FMR gas, the leading deviation of the pressure from the Stefan-Boltzmann law is linear in $T$.  However this linear term --- and only a linear term --- is annihilated in the formula for the energy density
$\e = T^2 { \p \over \p T} ( { p \over  T } )$, which gives
\beq
\label{asympte}
\e = 3 c_{SB} \ T^4 + O(1).
\eeq
An EOS of this type was obtained as a fit to the lattice data at high temperature in \cite{Kallman84}.  
For the speed of sound one obtains
\begin{equation}
c_s^2 = {\p p \over \p \e} 
= {1 \over 3} \Big(1 - { L \over 12c_{SB}T^3} \Big) 
+ O\Big( {1 \over T^4} \Big).
\end{equation}

\section{Comparison with numerical EOS}


The EOS of the FMR gas at high $T$ is not sensitive to the exact form of $E(k)$ because \rf{asymptp} and \rf{asympte} hold as long as $\g > 3$ and 
$E(k) \geq k$.  To compare with the numerical data, we take the Gribov dispersion relation \rf{dispersion}.  The unknown mass scale $M$ is determined by fitting the anomaly $\th = \e - 3p$ at high $T$, because the corrections to this quantity are expected to be small.  [Indeed, taking thermal perturbation theory as a guide~\cite{Kapusta}, we note that
the leading correction to $p$ is of order
$\D p \sim g^2(T) T^4 \sim { T^4 \over \ln T } $.  The anomaly is given by
$\th = T^5 { \p \over \p T }( { p \over T^4} )$, so the corresponding correction to the anomaly,
$\D \th = T^5 { \p \over \p T }( { \D p \over T^4} )$, is of order
$T^5 { \p \over \p T }( { 1 \over \ln T} ) 
= -   { T^4 \over \ln^2 T}  \sim g^4(T) T^4$.] 

\begin{figure}[htb!]
\includegraphics[width=3.3in]{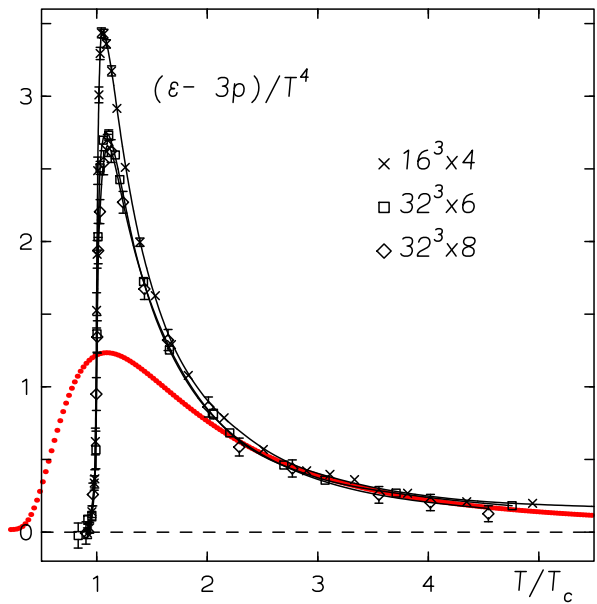}
\vspace{-0.5cm}
\caption{Numerical and analytic plot of  $(\e - 3p)/T^4$.}
\label{fig:pressure}
\end{figure}

\begin{figure}[htb!]
\includegraphics[angle=-90,width=3.3in]{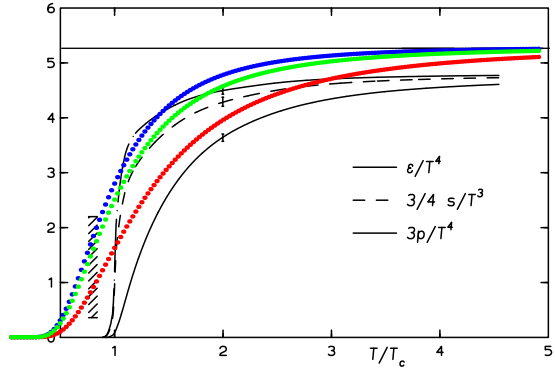}
\vspace{-2cm}
\caption{Numerical and analytic plots of $\e/T^4, \ 3/4 \ s/T^3, \ 3p/T^4$.}
\label{fig:pressure}
\end{figure}

The numerical data of \cite{Karsch96} for ${\th \over T^4}$ are represented by the black interpolating curves in Fig.~1. (More recent studies include dynamical fermions or a chemical potential that cannot be described by the FMR gas.) 
The data for $N_\t = 6$ and 8 agree, and were interpreted 
as continuum values~\cite{Karsch96}.  The red dots are obtained from the analytic formula \rf{anomalyx}, with mass scale set at $M = 2.6 T_c$ by fitting at high $T$, where $T_c$ is the transition temperature.  The relatively large difference 
in the transition region between the FMR gas and the numerical data for $\th = \e - 3p$ occurs because the FMR gas does not exhibit a sharp phase transition, whereas for pure SU(3) gauge theory there is a first order phase change, so $\e$ is discontinuous while $p$ is continuous.  We do not attempt to estimate the error of $M$ because perturbative-type corrections to the FMR gas have been neglected.  From the value $T_c = 0.625 \sqrt{\s}$ of \cite{Karsch96}, where $\s$ is the string tension, one gets 
$M = 1.6  \sqrt{\s}$, or $M = 705$ MeV, where the string tension for the quarkless theory is defined to be $\sqrt{\s} = 440$ MeV.
 
The numerical data of \cite{Karsch96} for ${\e \over T^4}$, 
${3 s \over 4 T^4}$ and ${3 p \over T^4}$ are
are displayed as black interpolating curves in Fig.~2, where 
${ s \over T^3} = { \e + p \over T^4}$.  The horizontal line represents the Stefan-Boltzmann EOS.  
For the FMR gas, ${3 p \over T^4}$ and 
${ \e \over T^4}$ approach the Stefan-Boltzmann limit like 
${ 1 \over T^3 }$ and ${1 \over T^4}$ respectively,
being quite close to it at the highest temperature displayed,
$T = 5T_c$, whereas the gluon plasma approaches the Stefan-Boltzmann limit more slowly.  The difference between the 
FMR gas and the gluon plasma in the range $2T_c$ to $5T_c$ appears attributable to perturbative-type corrections of moderate size.  In standard thermal perturbation theory \cite{Kapusta} these are of leading order $g^2(T) \sim {1 \over \ln T}$, but diverge at order $g^6$,
whereas corrections to the FMR gas are expected to be calculable.

\section{Discussion}
   
   	Above the transition region, the EOS of the FMR gas gives a good description of the most prominent feature of the gluon plasma which is the rapid drop of the pressure compared to the energy from the Stefan-Boltzmann value, as $T$ decreases from infinity.  The linear asymptotic trace anomaly \rf{lata} provides a ready explanation for this, that holds for any quasi-particle model with $\g > 3$ and 
${ \p \over \p k} \Big({E(k) \over k }\Big) < 0$, although other fits are certainly not excluded.  The FMR gas is not exact even at high temperature because of perturbative-type corrections.  We expect however that they are calculable and of moderate size above the transition region. 
	
	The transition region is not so well described by the FMR gas.  There is no sharp phase transition because $p(T)$ is an analytic function.  Moreover the dependence on N [of SU(N)] is only through the coefficient 
$(N^2-1)$, whereas even the order of the phase change depends on $N$, being second order for SU(2) and first order for SU(3). 
An analysis of the phase transition based on center symmetry is given in~\cite{Pisarski00}.   However the dispersion relation 
$E(k) = \sqrt{k^2 + {M^4 \over k^2} }$
has a minimum energy $E_{\rm min} = \sqrt 2 M$, so the thermodynamic functions $\e$, $p$ and $s$ of the FMR gas are exponentially small for 
$T < \sqrt 2 M = 997$ MeV (for $M = 705$ MeV).  
The mass of the lightest glueball is of order 1 GeV, so the thermodynamic functions of the FMR gas are exponentially small where they are supposed to be.  Altogether the FMR gas, with a single parameter $M$ which is the mass scale, is competitive with cut-off phenomenological models~\cite{DeGrand87}.

It may be helpful to compare the FMR gas with the more perfected
phenomenological models~\cite{Engels89}.  The dispersion relation, $E(k)$, which reflects the restriction to the FMR,  is $T$-independent, and consequently the thermodynamic identities are automatically satisfied using Planck's formula.  By contrast the phenomenological dispersion relations $E(k, T)$ 
of~\cite{Engels89} are $T$-dependent.    When used in Planck's formula, the $T$-dependent dispersion relation requires the introduction of an additional ``background field" $B(T)$ to assure that the thermodynamic identities are satisfied.  The improved phenomenological models of~\cite{Engels89} are more precise than the FMR gas especially in the transition region, but they also require several parameters.  We have fit only the QCD mass scale $M$.  

	However the FMR gas is not intended to be a precise phenomenological model, but rather to provide a useful starting point, well founded in the principles of gauge theory, that allows calculable, moderate size corrections at high~$T$.  Athough it is defined by Gribov's dispersion relation of 1978, the FMR gas has two important properties that were later independently reinvented:  (i)~Its EOS closely resembles simple phenomenological models of the gluon plasma~\cite{DeGrand87}.  (ii)~The effective mass 
${ M^2 \over k }$ controls infrared divergences in higher order corrections so it is not necessary to introduce the magnetic mass $m \sim g^2 T$~\cite{Pisarski81} for this purpose.


\bigskip
%
%
\acknowledgments{I am grateful for valuable conversations and correspondence with Jeff Greensite, Andrei Gruzinov, Stefan Olejnik, Carlo Piccioni, Lorentz von Smekal and Mike Strickland.  This research is supported in part by the National Science Foundation, Grant No. PHY-0099393. }

%
%

\end{document}